\documentclass[12pt]{article}
\usepackage{latexsym}
\usepackage{amsmath}
\usepackage{amsfonts}
\usepackage{amssymb}

\usepackage{theorem}

\numberwithin{equation}{section}

\newtheorem{theorem}{Theorem}[section]

\newtheorem{corollary}[theorem]{Corollary}
\newtheorem{lemma}[theorem]{Lemma}

\newenvironment{proof}{\begin{trivlist}\item[]{\sc Proof:}\/}{%
\hfill\mbox{$\Box$}\end{trivlist}}

\newcommand{\half}{\frac{1}{2}}
\newcommand{\mean}[1]{\langle #1\rangle}
\newcommand{\qand}{\quad{\rm and}\quad}
\newcommand{\fhat}{\widehat{f}}
\newcommand{\norm}[1]{\Vert #1\Vert}
\newcommand{\order}[1]{O(#1)}
\newcommand{\barlan}{\bar{\lambda}^{(N)}}
\newcommand{\Seq}{S^{(\text{eq},T)}}
\newcommand{\Nlambda}{\lambda^{(N)}}
\newcommand{\dlambda}{\ell_d}
\newcommand{\vep}{\varepsilon}
\newcommand{\rmd}{{\rm d}}
\newcommand{\rme}{{\rm e}}
\newcommand{\set}[1]{\{#1\}}
\newcommand{\tr}{{\rm Tr}}
\newcommand{\JN}{{J}^{(N)}}
\newcommand{\vc}[1]{\boldsymbol{#1}}
\newcommand{\ci}{{\rm i}}
\newcommand{\re}{{\rm Re\,}}

\newcommand{\1}{I}
\newcommand{\Z}{{\mathbb Z}}
\newcommand{\R}{{\mathbb R}}
\newcommand{\C}{{\mathbb C\hspace{0.05 ex}}}

\begin{document}

\newcommand{\email}[1]{Electronic mail: \tt #1}

\newcommand{\emailjani}{\email{jlukkari@ma.tum.de}}
\newcommand{\addressjani}{Centre for Mathematical Sciences, %
Munich University of Technology, 85747 Garching, Germany.}

\newcommand{\emailjoel}{\email{lebowitz@math.rutgers.edu}}
\newcommand{\addressjoel}{Department of Mathematics, Rutgers University, %
Piscataway, New Jersey 08854, USA.}

\newcommand{\emailfederico}{\email{bonetto@math.gatech.edu}}
\newcommand{\addressfederico}{School of Mathematics, %
Georgia Institute of Technology, Atlanta, Georgia 30332, USA.}

\title{Fourier's Law for a Harmonic Crystal with Self-consistent
   Stochastic Reservoirs}
\author{Federico Bonetto\thanks{\addressfederico\ \emailfederico.},
   Joel L.\ Lebowitz\thanks{\addressjoel\ \emailjoel.},
   and Jani Lukkarinen\thanks{\addressjani\ \emailjani.} }

\maketitle

\vspace*{-1.5em}
\begin{center}
{\em Dedicated to Elliott Lieb on the occasion of his seventieth
birthday} 
\end{center}
\vspace*{0em}

\begin{abstract}
We consider a $d$-dimensional harmonic crystal in contact with a stochastic
Langevin type heat bath at each site. The temperatures of the ``exterior''
left and right heat baths are at specified values $T_L$ and $T_R$,
respectively, while the  temperatures of the ``interior'' baths are chosen
self-consistently so that there is no average  flux of energy between them
and the system in the  steady state.  We prove that  this requirement
uniquely fixes the temperatures and the self consistent system has a unique
steady state.   For the infinite system this state is one of local thermal
equilibrium.  The corresponding heat current  satisfies Fourier's law with
a finite positive thermal conductivity which can also be computed using the
Green-Kubo formula.  For the harmonic chain ($d=1$) the conductivity
agrees with the expression obtained by Bolsterli, Rich and Visscher in 1970
who first studied this model.  In the other limit, $d \gg 1$, the
stationary infinite volume heat conductivity behaves as $(\dlambda d)^{-1}$
where $\dlambda$ is the coupling to the intermediate reservoirs.  We also
analyze the effect of having a non-uniform distribution of the heat bath
couplings.  These results are proven rigorously by controlling the behavior
of the correlations in the thermodynamic limit.
\end{abstract}


\section{Introduction}

Our understanding of non-equilibrium systems is at the present time very
incomplete.  In particular, we still have no model Hamiltonian system for
which Fourier's law has been proven rigorously.  A review of the problems
and of the few known exact results related to Fourier's law is given in
\cite{bonetto00} and in \cite{lepri01} which also contains a survey of
recent numerical results.

Here we study, in a mathematically rigorous manner, the microscopic
structure of the stationary non-equilibrium state of a ``self-consistent
harmonic crystal'', a model introduced by Bolsterli, Rich and Visscher
(BRV) in \cite{bolrich70,rich75}. This is a $d$-dimensional system of
$N_1\times \cdots\times N_d$ oscillators whose time-evolution is given by a
combination of Hamiltonian and stochastic dynamics.  The Hamiltonian is
composed of harmonic nearest neighbor and  ``on-site'' potentials and the
stochastic part comes from coupling each particle in the chain to its own
heat bath.

We want to describe a situation where we have a temperature gradient in one
direction (the ``first'' with $N_1$ oscillators) while the temperature  is
uniform in the remaining $d-1$ directions, on which we impose periodic
boundary conditions. The temperatures of the heat baths of the end-point
particles in the first direction are fixed to given values $T_L$ and $T_R$,
while the temperatures of the interior heat baths are chosen
self-consistently by the requirement that there is  no energy flux, on
average,  between any such reservoir and the system in the steady state.
{}From a physical point of view, we may think of the interior heat
reservoirs as representing schematically the effect of degrees of freedom
not included in the Hamiltonian.

Using numerical studies and non-rigorous arguments, BRV found that in the
case $d=1$ ({\it chain}\/) the (kinetic) temperature  profile of the system
in its steady state is linear, with a heat flux proportional to $N^{-1}$
for large $N$.  This corresponds to the self-consistent system having a
finite, temperature independent, thermal conductivity
\cite{bonetto00,lepri01}.    These results are in sharp contrast to those
found earlier by Rieder, Lebowitz and Lieb \cite{lebo67}, who studied a
system with the same Hamiltonian dynamics, but with heat baths acting only
at the boundaries.  They found that the system had an infinite conductivity
and a constant temperature profile away from the ends, results later
generalized to the higher dimensional case by Nakazawa \cite{nakazawa70}.

In this paper, we provide a rigorous proof that the steady state of the
self-consistent system has indeed the properties found by BRV for the $d=1$
case in  \cite{bolrich70,rich75}, and we extend the results to cover all
$d\geq 1$.  More precisely, we show that, in the limit where all 
$N_i\to\infty$, the steady state is a local equilibrium state 
\cite{spohn:hydro} with a temperature profile satisfying Fourier's law with
a finite, temperature independent, thermal conductivity.

We  deal first with the $d=1$ case, and  consider the higher dimensional
case only in section \ref{sec:crystal}.   We define the model  and solve
its dynamics with a given temperature profile in section
\ref{sec:dynamics}.  We then turn to the self-consistency condition in
section \ref{sec:selfcons}, proving in particular, that the self-consistent
profile is always uniquely determined by the boundary temperatures.
Section \ref{sec:fourier} contains our main results: we prove there the
local equilibrium property and Fourier's law.  In section
\ref{sec:greenkubo}, we use the explicit solution to show that the
Green-Kubo formula holds for this system.  In section \ref{sec:gener}, we
briefly analyze the case where the couplings to the heat baths are
non-uniform, and we conclude that the local macroscopic heat conductivity
is proportional to the inverse of the local average of the couplings.

Finally in section \ref{sec:crystal}, we first show how one can map the
higher dimensional self-consistent system with periodic boundary condition 
on all directions but the first into a set of one-dimensional chains, and
then apply the earlier results to derive a generalization to the higher
dimensional case. We show there that for large $d$ the conductivity
behaves as $(\dlambda d)^{-1}$ where $\dlambda$ is the
coupling to the intermediate 
reservoirs.  Some technical details of the calculations
are collected in the appendices.

\section{Dynamics and the stationary state}
\label{sec:dynamics}

To start with, we consider a chain of $N$ oscillators with a Hamiltonian 
\begin{equation}
  \label{eq:hamiltonian}
  H(\vc{q},\vc{p}) = \sum_{i=1}^N \Bigl[\frac{1}{2} p_i^2 + u(q_i)\Bigr] 
  + \sum_{i=1}^{N+1} v(q_i-q_{i-1}),
\end{equation}
where $\vc{q}$ and  $\vc{p}$ are vectors
in $\R^N$, we set $q_0=0=q_{N+1}$, and we have
\begin{equation}
  \label{eq:defUV}
  u(q) = \frac{1}{2} \gamma^2 q^2 \qand v(q) = \frac{1}{2} \omega^2 q^2
\end{equation}
with $\gamma,\,\omega>0$.  In addition, the oscillator at each site $i$ is
coupled to a Langevin heat bath at temperature $T_i \ge 0$, with a coupling
strength $\lambda>0$.  As in \cite{lebo67}, the time-evolution of the
system is then given by the stochastic differential equations,
\begin{equation}\label{e:OUprocess}
 \dot {\vc{X}} = -A \vc{X}\, + \Sigma \dot{\vc{W}}
\end{equation}
where $\vc{X} = (\vc{q},\vc{p})$ is the phase-space vector, the
$\dot{W}_i$ are independent white noises and $A$ and $\Sigma $ are $2N$ by
$2N$ matrices given by
\begin{equation}
  \label{eq:defA}
  A = \left( \begin{array}{cc} 0 & -\1 \\ 
     \Phi & \Lambda \end{array} \right)\qand
 \Sigma = \left( \begin{array}{cc} 0 & 0 \\ 
     0 & \sqrt{2\Lambda {\cal T}}
   \end{array} \right) . 
\end{equation}
Here $\1$ is the unit $N$ by $N$ matrix, $\Lambda_{ij}= \delta_{ij}
\lambda$, ${\cal T}_{ij}=\delta_{ij} T_i$, and
\begin{equation}
  \label{eq:defphi}
  \Phi = \omega^2 (-\Delta + \nu^2 I),
\end{equation}
where $\nu^2 = \gamma^2/\omega^2$ and $\Delta$ denotes the discrete
Laplacian with Dirichlet boundary conditions:
\[
\Delta_{ij} = -2 \delta_{i,j} + \delta_{i-1,j} + \delta_{i+1,j}.
\]

Equations (\ref{e:OUprocess}) define an Ornstein-Uhlenbeck
process whose solution with initial data $\vc{X}(0)$ is given
by the stochastic integral (for details, see e.g.\ chapter 5 in
\cite{oksendal})
\begin{equation}
  \label{eq:Xevint}
  \vc{X}(t)= \rme^{-t A}\vc{X}(0) + \int_{0}^t \!\rmd s\, 
    \rme^{-(t-s) A}\Sigma \dot{\vc{W}}(s).
\end{equation}

This is a Gaussian process, determined uniquely by its mean and
covariance which can be computed directly from (\ref{eq:Xevint}).
However, for latter use we assume now that $\vc{X}(0)$ is distributed
according to a Gaussian measure with a mean $\vc{X}_0$ and a
covariance $C_0$---the deterministic case is then obtained by setting
$C_0=0$.  Then the mean evolves by
\begin{equation}
  \label{eq:Xmean}
  \mean{\vc{X}(t)} = \rme^{-t A}\vc{X}_0,
\end{equation}
and for the covariance $C(t',t) \equiv
  \mean{[\vc{X}(t'){-}\mean{\vc{X}(t')}] \otimes
  [\vc{X}(t){-}\mean{\vc{X}(t)}]}$ we get
\begin{align}
  \label{eq:ctpt}
  C(t',t) & = \left\{ \begin{array}{ll} 
      \rme^{-(t'-t) A}\, C(t,t) & \text{if }t' \ge t \\ 
      C(t',t')\,\rme^{-(t-t') A^T} & \text{if }t' \le t
    \end{array}\right. ,  \\
\intertext{with}
  \label{eq:Cteq}
  C(t,t) & =  \rme^{-t A} C_0 \,\rme^{-t A^T} +
        \int_{0}^{t}\!\rmd s \, \rme^{-s A}\Sigma^2\rme^{-s A^T}.
\end{align}

We show in Appendix \ref{sec:boundexpA} that for any $\alpha$
satisfying 
\[
  0<\alpha <
  \min\Bigl\{ \frac{\lambda}{2}, \frac{\gamma^2}{\lambda} \Bigr\}
\]
we can find a constant $c<\infty$ such that
for all $t>0$ and for all $N$
\begin{equation}
  \label{eq:exptAbound}
  \norm{\rme^{-t A}} \le 
  c \,\rme^{-t\alpha }. 
\end{equation}
The uniform exponential decay of $\rme^{-t A}$ implies that there is
an exponentially fast convergence in the microscopic scale
to a unique stationary state, which
is Gaussian with mean $\vc{0}$ and covariance $S$,
\begin{gather}\label{e:defstationary}
  S = \int_{0}^{\infty}\!\rmd s \, \rme^{-s A}\Sigma^2\rme^{-s A^T}.
\end{gather}

This $S$ is the unique solution of the equation
\begin{gather}\label{e:statstate}
  A S + S A^T = \Sigma^2,
\end{gather}
which we solve following ref.\ \cite{bolrich70}. We
divide $S$ into $N$ by $N$ components,
\[
  S = \left( \begin{array}{cc} 
      U & Z \\ Z^T & V
 \end{array} \right),
\]
and get the following four equations equivalent to
(\ref{e:statstate}):
\begin{align}\label{eq:sseqns2}
\begin{split}
  Z = -Z^T, \quad
  V & = \frac{1}{2} (\Phi\, U + U \Phi ) + 
  \frac{1}{2} ( Z\Lambda - \Lambda Z ), \\ 
 Z\Lambda +  \Lambda Z & = \Phi\, U - U \Phi, \\  
 \Lambda ( {\cal T} - V)+ ( {\cal T} - V) \Lambda 
 & = \Phi Z - Z \Phi.
\end{split} 
\end{align}

A diagonalization of the discrete Laplacian yields
\[
  \Phi = FY\!F^T,
\]
where $Y_{kl} = \mu_k \delta_{kl}$, $\mu_k$ are the eigenvalues of $\Phi$:
\begin{equation}
  \label{eq:defmuk}
  \frac{ \mu_k }{\omega^2}
  = \nu^2 + 4 \sin^2\Bigl(\frac{\pi k}{2(N+1)}\Bigr),
\end{equation}
and $F$ is the orthonormal matrix
\begin{equation}
  \label{eq:defF}
  F_{kl} = \sqrt{\frac{2}{N+1}} \sin\Bigl(\frac{\pi kl}{N+1}\Bigr).
\end{equation}
As shown in Appendix \ref{sec:statstate}, any block $B$ of the
covariance matrix (i.e.\ $B=U$, $V$ or $Z$) can then be obtained by a
linear transformation of the form
\begin{equation}
  \label{eq:Bij}
  B_{ij} = \sum_{r=1}^N B_{ij}^{(r)} T_r
\end{equation}
where
\begin{gather}\label{e:covM}
   B_{ij}^{(r)} = 
   \sum_{k,l=1}^N F_{ik} F_{jl} f^{(B)}(c_k,c_l) F_{rk} F_{rl}
\end{gather}
and
\begin{equation}
\label{eq:defck}
  c_k = \cos\left(\frac{\pi k}{N+1}\right).
\end{equation}
The functions $f^{(B)}$ for the different choices of $B$ are given by
\begin{align}\label{eq:deffX}
\begin{split}
  f^{(U)}(x,y) & = \frac{\lambda^2}{\omega^4} \frac{1}{G(x,y)} \\ 
  f^{(V)}(x,y) & = 1- \frac{(x-y)^2}{G(x,y)} \\ 
  f^{(Z)}(x,y) & = \frac{\lambda}{\omega^2} \frac{y-x}{G(x,y)} 
\end{split}
\end{align}
where
\begin{equation}
  \label{eq:defG}
  G(x,y) = (x-y)^2 +  \frac{\lambda^2}{\omega^2} ( \nu^2 + 2 - x -y ).  
\end{equation}

When $T_i=T$ for all $i$, only the values with $k=l$ contribute in
equation (\ref{e:covM}).  The stationary covariance is then given by
\begin{equation}
  \label{eq:eqlcov}
  \Seq  = T \left( \begin{array}{cc} 
      \Phi^{-1} & 0 \\ 0 & \1
 \end{array} \right),
\end{equation}
and the stationary measure is the Gibbs measure at temperature $T$.

\subsection{Energy current} 
\label{sec:current}

We define the local energy of particle $i$ by
\[
  H_i(\vc{q},\vc{p}) = \frac{1}{2} p_i^2 + u(q_i) + 
    \frac{1}{2} \left[v(q_i-q_{i-1})+v(q_{i+1}-q_{i})\right]
\]
for $i=2,\ldots,N-1$.  The boundary terms ($i=1,N$) are defined similarly,
but with double the interaction energy contribution from the connections
to $q_0=0$ and to $q_{N+1}=0$.  With these
definitions,
\[
  H(\vc{q},\vc{p}) = \sum_{i=1}^N H_i(\vc{q},\vc{p})
\]
and 
\begin{equation}
  \label{eq:energyflux}
  \frac{\rmd}{\rmd t} \langle H_i(t) \rangle = 
  -\left[\langle J_i(t)\rangle - 
    \langle J_{i-1}(t) \rangle \right] + \mean{R_i(t)},
\end{equation}
where
\begin{equation}
  \label{eq:Jidef}
J_i(\vc{q},\vc{p}) = -\frac{\omega^2}{2} (q_{i+1}-q_i)
     \left( p_i + p_{i+1}\right),  \quad i=1,\ldots,N-1,
\end{equation}
$J_i=0$ for $i=0,N$, and 
\[
  R_i(\vc{q},\vc{p}) = \lambda \left( T_i - p_i^2 \right).
\]
The $J_i$ correspond to the energy currents inside the system while
$R_i$ gives the energy flux from
the $i$-th reservoir to the $i$-th oscillator.

The corresponding expectation values in the stationary state are
\begin{equation}\label{eq:ZJrel}
  \langle J_i \rangle_S = 
    \frac{\omega^2}{2} \Bigl(-Z_{i+1,i} -Z_{i+1,i+1}
    + Z_{i,i} + Z_{i,i+1} \Bigr) = \omega^2 Z_{i,i+1},
\end{equation}
where we have used the antisymmetry of $Z$, and
\begin{equation}
  \label{eq:defensources}
    \langle R_i \rangle_S = \lambda \left( T_i - V_{ii} \right).
\end{equation}

\section{Self-consistency condition}
\label{sec:selfcons}


As described in the introduction, we let the end-point temperatures 
$T_1 = T_L$ and $T_N = T_R$ be given independently of $N$.  
Then we want to choose
$T_i$ for $i=2,\ldots,N-1$ in such a way that $\langle R_i \rangle_S = 0$
or, by (\ref{eq:defensources}), so that
\begin{equation}
  \label{eq:sccond}
  T_i = V_{ii}.
\end{equation}

By (\ref{e:defstationary}), the kinetic temperature vector
$(V_{ii})_{i=1}^N$ depends linearly on the imposed temperature vector
$\vc{T}$.  Therefore, all solutions to the self-consistency condition can
be obtained using the equation
\begin{equation}
  \label{eq:tsclin}
  T_i = T_R + (T_L-T_R) T^{(1,0)}_i
\end{equation}
where $\vc{T}^{(1,0)}$ is a solution of the problem with $T_L=1$
and $T_R=0$.  The set of all such $\vc{T}^{(1,0)}$ form a convex set, i.e.\
any non-uniqueness in the solution of 
the self-consistency condition  would imply the existence of a whole
continuum of solutions.  We shall now prove that for our model the solution
to the self-consistency problem is unique.

Let us denote the above linear mapping from imposed to kinetic temperatures
by $M$, i.e.\ $V_{ii} = \sum_j M_{ij} T_j$.  It follows straightforwardly
from   (\ref{e:defstationary}) that 
$M_{ij}\ge 0$ for all pairs $i,j$, and by the explicit solution given in
(\ref{eq:Bij})--(\ref{eq:defG}) we have
\begin{equation}
  \label{eq:defM}
  M_{ij} = V_{ii}^{(j)} =  \sum_{k,l=1}^N F_{ik} F_{il} f_{kl} F_{jk}
    F_{jl},
\end{equation}
where
\[
  f_{kl} = f^{(V)}(c_k,c_l) = 1 -\frac{(c_k-c_l)^2}{G(c_k,c_l)}\ge 0.
\]
By (\ref{eq:defM}), $M$ is then symmetric and satisfies for all $i$
\begin{equation}
  \label{eq:sumofm}  \sum_{j=1}^N M_{ij} = \sum_{j=1}^N M_{ji} = 1,
\end{equation}
which imply that $M$ is, in fact, a doubly stochastic matrix.

\begin{theorem}\label{th:scprof}
For any given end-point temperatures $T_L$, $T_R\ge 0$, there is a unique,
positive temperature profile which satisfies the self-consistency
condition.  In addition, all temperatures in the profile lie between the
end-point temperatures.
\end{theorem}
\begin{proof}
Let $\vc{T}$ be a self-consistent profile and define
\[
  a_i = \left\{ \begin{array}{ll}
      T_i-T_R, & \text{for }i=2,\ldots,N-1, \\
      0, & \text{if }i=1\text{ or } i=N
    \end{array} \right. .
\]
Clearly $\vc{a}$ belongs to the subspace
\[
  W = \Bigl\{\vc{x}\in\R^N \, \Big|\, x_1=0 = x_N \Bigr\}.
\]
Let us denote the orthogonal projection to
the subspace $W$ by $P_W$, and
define $Q=P_W M P_W$. Using (\ref{eq:sumofm}) in (\ref{eq:sccond}) we
get the equation
\begin{equation}
  \label{eq:truesc}
  \vc{a} = Q \vc{a} + \vc{b}
\end{equation}
where the vector $\vc{b}$ is defined by
\[
  b_i = \left\{ \begin{array}{ll}
      M_{i1} (T_L-T_R) , & \text{for }i=2,\ldots,N-1, \\
      0, & \text{if }i=1\text{ or } i=N
    \end{array} \right. .
\]

We shall later prove in Corollary \ref{th:Qnorm} that $\norm{Q} < 1$.
Then for any $\vc{b}\in \R^N$, equation (\ref{eq:truesc}) has a unique
solution given by
\[
  \vc{a} = \sum_{n=0}^\infty Q^n \vc{b},
\]
and the self-consistent profile must thus satisfy
\begin{equation}
  \label{eq:Tdef}
  T_i = T_R + \sum_{n=0}^\infty (Q^n \vc{b})_i,
  \quad\text{for } i = 2,\ldots,N-1.
\end{equation}

As the profile defined by (\ref{eq:Tdef}) also satisfies (\ref{eq:sccond}),
this proves the existence and uniqueness of the self-consistent
profile.  Repeating the above computation for $\vc{a}'=P_W(\vc{T}-T_L
\vc{1})$
instead of $\vc{a}$, and then using the positivity of
$M_{ij}$ in the equations corresponding to (\ref{eq:Tdef}), proves
that $T_R\le T_i\le T_L$ for all $i$ when $T_R \le T_L$,
and that $T_L\le T_i\le T_R$ when $T_R \ge T_L$.
\end{proof}

To conclude the proof,  we still need to prove that $\norm{Q} < 1$.  This
will follow from the following lemma:
\begin{lemma} \label{th:Qlowerbound}
Let $c$ be a constant which satisfies
\[
  G(x,y) \le \frac{c}{2},
\]
for all $x,y \in [-1,1]$.  Then, for any vector $\vc{x}\in \R^N$,
\begin{equation}
  \label{eq:1Mbound}
  \vc{x}^T\! (\1-M)\, \vc{x} \ge \frac{1}{c} \norm{D \vc{x}}^2,  
\end{equation}
where $D$ is the ``finite difference operator'' defined by
\[
  (D \vc{x})_i = \left\{ \begin{array}{ll}  
      x_i-x_{i+1}, & \text{for } 1 \le i \le N-1 \\ 
      0, & \text{for } i = N
    \end{array} \right. .
\]
\end{lemma}
\begin{proof}
Let $\vc{x}\in \R^N$, and define
$\tilde{x}_{kl} = \sum_{i=1}^N F_{ik} F_{il} x_i$
for all $k,l$.  Then by (\ref{eq:defM}) and the assumption
$G(c_k,c_l) \le c/2$,
\begin{align*}
  \vc{x}^T\! (\1-M)\, \vc{x} &
  \ge \frac{2}{c} \sum_{k,l=1}^N \tilde{x}_{kl}^2 (c_k-c_l)^2 \\
  &= \frac{2}{c} \sum_{i,j=1}^N x_i x_j \sum_{k,l=1}^N
  F_{ik} F_{il} F_{jk} F_{jl} ((1-c_k)-(1-c_l))^2.
\end{align*}
But by an explicit computation, we have for all $i,j$,
\[
  2 \sum_{k,l=1}^N F_{ik} F_{il} F_{jk} F_{jl} ((1-c_k)-(1-c_l))^2 =
    (\Delta^2)_{ij} \delta_{ij} - (\Delta_{ij})^2 = (D^T\! D)_{ij},
\]
and the inequality (\ref{eq:1Mbound}) has been derived.
\end{proof}
For the applications, we  note that a constant $c$ satisfying the
requirement of the lemma  can always be found.
\begin{corollary}\label{th:Qnorm}
 $\norm{Q} < 1$.
\end{corollary}
\begin{proof}
Since $W$ is finite-dimensional and $Q=P_W Q P_W$ is symmetric, it will be
enough to show that for all  $\vc{x}\in W$ with $\norm{\vc{x}}\le 1$
\begin{equation}
  \label{eq:xQx}
 0\le \vc{x}^T\! Q \vc{x} < 1.
\end{equation}
But for such $\vc{x}$, $\vc{x}^T\! Q \vc{x} = \vc{x}^T\! M \vc{x}$, and the
lemma together with $f_{kl}\ge 0$ yields
\begin{equation}
  \label{eq:xQx2}
  0\le \vc{x}^T\! M \vc{x} \le 1 - \vc{x}^T (\1-M) \vc{x} \le 1-
   \frac{1}{c} \norm{D \vc{x}}^2.
\end{equation}
If $D \vc{x} = \vc{0}$, then $x_i = x_1 = 0$ for all $i$, and we must now
have either $\norm{D \vc{x}}>0$, when (\ref{eq:xQx2}) implies
(\ref{eq:xQx}), or $\vc{x}=0$, when (\ref{eq:xQx}) is trivially true.
\end{proof} 

\section{Fourier's law}
\label{sec:fourier}

In this section we first derive a number of technical estimates which will
be necessary to control the behavior of the system in the thermodynamic
limit.  We then show that the self-consistent  steady state is
microscopically a local equilibrium state with a heat flux satisfying
Fourier's law.

\subsection{Exponential decay of correlations}
\label{sec:correxpdec} 

Let $f$ denote any one of the three functions $f^{(B)}$ defined by equation
(\ref{eq:deffX}).  Clearly, $f$ is a rational function, analytic everywhere
but at the zeroes of  $G$ in (\ref{eq:defG}).  On the other hand,
\begin{equation}
\label{5.1}
  G(x,y) \ge \frac{\lambda^2\nu^2}{\omega^2} = \lambda^2
  \frac{\gamma^2}{\omega^4}
\end{equation}
for all $x$, $y\in [-1,1]$, and since we have assumed that $\gamma>0$,
there are no zeroes of $G$ inside $[-1,1]^2$.

This implies that the function $f(\cos\cdot, \cos\cdot)$, which enters in
(\ref{e:covM}), is analytic in some neighborhood region of $[-\pi,\pi]^2$
in $\C^2$.  In particular, its Fourier series converges pointwise, and we
have for all $x$, $y \in \R$,
\begin{equation}
\label{e:deffhat}
   f(\cos x,\cos y) = \sum_{m,n= -\infty}^{\infty} \cos(m x) \cos(n y)
   \fhat(m,n).
\end{equation}

Applying this with $f=f^{(B)}$ in (\ref{e:covM})  we get after some
straightforward algebra,
\begin{equation}
  \label{eq:Bfhat}
  B_{ij}^{(r)} = \fhat_N(i-r,j-r) + \fhat_N(i+r,j+r) - \fhat_N(i-r,j+r) -
  \fhat_N(i+r,j-r)
\end{equation}
where $\fhat_N$ is defined by
\begin{equation}
  \label{eq:fhatN}
   \fhat_N(m,n) = \sum_{k,l\in\Z} \fhat(m+2(N{+}1) k,n+2(N{+}1) l).
\end{equation}

By the above mentioned analyticity, the Fourier coefficients $\fhat$ decay
exponentially.  From this the following behavior for $\fhat_N$ is easily
derived:
\begin{lemma}\label{th:expdecay}
For any block $B$, define $\fhat_N$ by (\ref{eq:fhatN}) using
$f=f^{(B)}$. Then there are strictly positive, $N$-independent constants
$a$ and $\alpha$ such that, for any $m,n\in\Z$,
\begin{equation}
  \label{eq:fNdecay}
  |\fhat_N(m,n)| \le a \, \rme^{-\alpha (|m'|+|n'|)}
\end{equation}
where $m' = (m \bmod 2(N+1)) \in \set{-N,\ldots,N+1}$ and $n$ defines $n'$
similarly.
\end{lemma}

\subsection{Local equilibrium}
\label{sec:localeql}

Consider some temperature profile $\vc{T} = (T_i)_{i=1}^{N}$, and let  its
maximum nearest neighbor variation be denoted by $\vep_N$, i.e.\ with $D$
defined as in Lemma \ref{th:Qlowerbound}, let
\begin{equation}\label{eq:grad}
  \vep_N = \max_i |(D \vc{T})_i|=\max_{i<N}|T_i-T_{i+1}|.
\end{equation}

We shall prove in this section that, if $\vep_N\ll 1$ and $\max_i T_i$ is
bounded, the local microscopic properties of the stationary measure near a
site $i$ can be well approximated by using the equilibrium measure with the
temperature $T_i$.  This will, in particular, justify our identification of
the parameter $T_i$ as a local temperature.

The main ingredient of the proof is the following corollary to Lemma
\ref{sec:correxpdec}:
\begin{corollary}\label{th:Bbound}
For any block $B$, there is an $N$-independent constant $a'$, such that for
all $i,j\in\set{1,\ldots,N}$,
\begin{equation}
  \label{eq:mateldecay}
  \sum_{r=1}^N |i-r| \, \bigl|B_{ij}^{(r)}\bigr| \le a'.
\end{equation}
\end{corollary}
\begin{proof}
Apply equation (\ref{eq:Bfhat}), the triangle inequality, and Lemma 
\ref{th:expdecay} to replace $|B_{ij}^{(r)}|$ in (\ref{eq:mateldecay}) by 
four exponential terms. The estimate follows  straightforwardly.
\end{proof}

Let  $B$ be any block of the stationary covariance matrix, and let $i$,
$j\in \set{1,\ldots,N}$.  As it takes $|r-i|$ ``steps'' to get from a site
$r$ to the site $i$, we have the obvious bound
\begin{equation}
  \label{eq:boundvarT}
  |T_r - T_i| \le |r-i| \vep_N.
\end{equation}
Then Corollary \ref{th:Bbound} immediately yields the estimate
\begin{equation}
  \label{eq:XminusXeq}
  \Bigl|B_{ij} - T_i \sum_{r=1}^N B^{(r)}_{ij}\Bigr| \le a' \vep_N.
\end{equation}

The value of $T_i \sum_{r=1}^N B^{(r)}_{ij}$ gives the component of the
stationary covariance matrix when all the temperatures are set equal to
$T_i$, i.e.\ the equilibrium covariance $B^{({\rm eq}, T_i)}$ at
temperature $T_i$, see (\ref{eq:eqlcov}).  As $a'$ in (\ref{eq:XminusXeq})
can be chosen independently of $N$, we have now shown that all
pair-correlations satisfy
\begin{equation}
  \label{eq:BBeq}
  B_{ij} = B^{({\rm eq}, T_i)}_{ij} + \order{\vep_N}.
\end{equation}

As both the equilibrium measure and the above stationary measure are
Gaussian, we can also conclude that all finite correlations  can be
approximated by the local equilibrium values, with an error which vanishes
when $\vep_N\to 0$.

\subsection{Self-consistent current and Fourier's law}

Let $\vc{T}$ now denote the self-consistent profile which by Theorem
\ref{th:scprof} is unique and is bounded by $T_L$ and $T_R$.  Then by
(\ref{eq:sseqns2}) and (\ref{eq:ZJrel}) we get for all $i=2,\ldots,N-1$,
\[
 0 = \lambda ({\cal T}- V)_{ii} = \frac{1}{2} (\Phi Z - Z \Phi)_{ii}  =
 \omega^2 (-Z_{i-1,i}-Z_{i+1,i}) = \mean{J_{i}}_S - \mean{J_{i-1}}_S.
\]
Therefore, the steady state current is constant throughout the chain:
\begin{equation}
  \label{eq:constcurrent}
  \mean{J_i}_S = \JN, \quad \text{for }i=1,\ldots,N-1.
\end{equation}

The same equations also imply the following relation between the local
steady state current and the $q$-$q$ correlations: for all $i=1,\ldots,N-1$,
\begin{align}  \label{eq:jJrel}
\begin{split} 
  \mean{J_i}_S & 
  = \omega^2 Z_{i,i+1} = \frac{\omega^2}{2 \lambda} 
  (\Phi\, U-U\Phi)_{i,i+1}\\  
  & = \frac{\omega^4}{2 \lambda} 
  (U_{ii} + U_{i,i+2} - U_{i-1,i+1} -  U_{i+1,i+1} ),
\end{split} 
\end{align}
where we define $U_{ij}=0$ when $i$ or $j\not\in\set{1,\ldots,N}$.  Summing
(\ref{eq:jJrel}) over all the indices $i$, we get by using equation
(\ref{eq:constcurrent})
\begin{equation}
  \label{eq:juu}
  (N-1) \JN = \frac{\omega^4}{2\lambda} (U_{11} - U_{NN}).
\end{equation}

Let then $D$ and $c$ be given as in Lemma
\ref{th:Qlowerbound}, and as before let 
$\vep_N = \max_i |(D \vc{T})_i|$.  Then the magnitude of 
the current and $\vep_N$ are related by the formula
\begin{equation}
  \label{eq:boundMN}
 \vep_N \le \sqrt{ \frac{c}{\lambda} ( T_L-T_R ) \JN }.
\end{equation}

To see this, first let $\vc{x} = (\1-M)\vc{T}$ where the matrix $M$ was
defined in section  \ref{sec:selfcons}.  Since $\vc{T}$ is self-consistent,
$x_i=0$ except possibly at the end-points.  But then by (\ref{eq:sumofm}),
$x_1+x_N=\sum_i x_i =0$, and thus also $x_N=-x_1$.  Therefore,
(\ref{eq:boundMN}) follows from Lemma \ref{th:Qlowerbound}, since then
\[
  x_1 ( T_L-T_R ) = \vc{T}^T (\1-M)\vc{T} \ge  \frac{1}{c} \norm{D
  \vc{T}}^2 \ge \frac{1}{c} \vep_N^2,
\]
and, by (\ref{eq:sseqns2}), $x_1 = T_1 - V_{11} = 
\frac{1}{2\lambda} (\Phi Z-Z\Phi)_{11} = \mean{J_1}/\lambda$.

Using (\ref{eq:boundMN})
we can estimate the error made when the terms on the right side of 
(\ref{eq:juu}) are replaced by their equilibrium values.  Equation
(\ref{eq:XminusXeq}) and the explicit form of the equilibrium
covariance given in (\ref{eq:eqlcov}) yield 
\begin{equation}
  \label{eq:JNbound}
 \left|(N-1) \JN - \frac{\omega^4}{2\lambda} 
   (\Phi^{-1}_{11} T_L  - \Phi^{-1}_{NN}T_R) \right| 
 \le a' \frac{\omega^2}{\lambda} \sqrt{ \frac{c}{\lambda} ( T_L-T_R ) \JN }
\end{equation}
where $a'$ is chosen as in Corollary \ref{th:Bbound} for the block $B=U$.

It follows from symmetry that 
$\Phi^{-1}_{NN}=\Phi^{-1}_{11}$, which has the limit 
\begin{equation}
  \label{5.16}
  \lim_{N\to\infty} \Phi^{-1}_{11} =  
  \frac{2}{\omega^2} \int_0^1\!\!\rmd x \, 
    \frac{\sin^2(\pi x)}{ \nu^2+ 4 \sin^2\bigl(\frac{\pi x}{2}\bigr)} 
    = \frac{2}{\omega^2} \frac{1}{2+\nu^2+\sqrt{\nu^2 ( 4+\nu^2 )} }.
\end{equation} 
It is then a consequence of (\ref{eq:JNbound}) that
\begin{equation}
  \label{5.18}
  \lim_{N\to\infty} (N-1) \JN = \kappa(T_L-T_R)
\end{equation}
where
\begin{equation}
  \label{eq:defkappa}
  \kappa = \frac{\omega^2}{\lambda} 
  \frac{1}{2+\nu^2+\sqrt{\nu^2 ( 4+\nu^2 )} }.
\end{equation}

Since $\JN = \order{N^{-1}}$, we get from (\ref{eq:boundMN}) that
\begin{equation}
  \label{5.19}
 \vep_N = \order{N^{-\frac{1}{2}}}.
\end{equation}
By our discussion in section \ref{sec:localeql}, this implies that in the
limit $N\to\infty$, all the correlation  functions involving finitely many
terms will converge to the corresponding local equilibrium values if we
identify $T_i$ with the local temperature of the system at a site $i$.

Summing (\ref{eq:jJrel}) from 1 to $j-1$ and combining it with
(\ref{eq:juu}) yields 
\begin{equation}
  \label{5.20}
  \frac{j-1}{N-1} (U_{11}-U_{NN}) = U_{11}- U_{jj}+ U_{j-1,j+1}.
\end{equation}
Then the local equilibrium approximation, equation (\ref{eq:BBeq}), 
shows that
\begin{equation}
  \label{5.21}
  \frac{j-1}{N-1} \Phi^{-1}_{11} (T_L-T_R) = T_L \Phi^{-1}_{11} - T_j
  \Phi^{-1}_{jj} + T_{j-1}\Phi^{-1}_{j-1,j+1} + \order{\vep_N}.
\end{equation}
For all $j$ and $k$,
$F_{j-1,k} F_{j+1,k} = F_{jk}^2 - F_{1k}^2$, and $\Phi^{-1}$ thus
satisfies the identity
\begin{equation}
  \label{5.22}
  \Phi^{-1}_{j-1,j+1} = \Phi^{-1}_{jj} - \Phi^{-1}_{11} .
\end{equation}
Since $|\Phi^{-1}_{jj}| \le 1/\gamma^2$ and $\Phi^{-1}_{11}>0$
uniformly in $N$, we then get the result
\begin{equation}
  \label{5.23}
  T_j = T_L + \frac{j-1}{N-1} (T_R-T_L) + \order{\vep_N}
\end{equation}
where the correction term vanishes uniformly in $j$ when $N\to\infty$.  

Setting $x = j/N$ the system therefore approaches,  in the limit
$N\to\infty$, a local equilibrium state with a temperature profile
\begin{equation}
  \label{5/24}
  T(x) = T_L + x (T_R-T_L), \quad x \in [0,1].
\end{equation}
Thus Fourier's law holds for the steady state of the system, and the
thermal conductivity  is given by the temperature independent constant
$\kappa$ in (\ref{eq:defkappa}).  Note that $\kappa$ remains finite when
$\nu\to 0$, which points towards a finite conductivity even for the system
without the on-site binding potential.

Let us finally remark that we do not think the above bound for $\vep_N$ is
optimal.  Preliminary numerical simulations suggest that
$\vep_N=\order{N^{-1}}$ rather than $\order{N^{-\half}}$.

\section{The Green-Kubo formula}
\label{sec:greenkubo}

The Green-Kubo formula expresses the local equilibrium conductivity at a
position $\vc{x}$ with temperature $T(\vc{x})$ as an integral over the
current-current correlations in a (closed) equilibrium system at uniform
temperature $T = T(\vc{x})$. This corresponds, for the type of stationary
state we consider, to a formula for $\kappa$ when $T_L$ and $T_R \to T$.
It is not immediately apparent how the presently available derivations of
such a formula (for recent results, see e.g.\ \cite{lebspohn99,eyink96})
could be applied to a stochastic system like the one considered here.  In
particular, it is not clear {\em which\/} current we should use in the
formula: i.e.\ how to include the stochastic source terms in
(\ref{eq:energyflux}).

In this section, we shall make an explicit computation which shows that the
form of the Green-Kubo formula, as defined e.g.\ in \cite{lepri01}, leads
to the correct conductivity for the system with the non-zero on-site
potential.

\begin{theorem}\label{th:GK}
Given $T>0$,
\begin{equation}
  \label{eq:defKG}
 \kappa_{\rm GK}(T) = \frac{1}{T^2} \int_0^\infty\!\rmd t \lim_{N\to\infty}
    C_{JJ}^{(N)}(t;T) = \lim_{N\to\infty} \frac{1}{T^2} 
      \int_0^\infty\!\rmd t\, C_{JJ}^{(N)}(t;T) = \kappa.
\end{equation}
\end{theorem}
In the theorem, $\kappa$ is defined by (\ref{eq:defkappa}), and
\begin{equation}\label{eq:defCJ}
  C_{JJ}^{(N)}(t;T) = \frac{1}{N+1} 
  \mean{J(\vc{q}(t),\vc{p}(t)) J(\vc{q}(0),\vc{p}(0))}_{S_T}
\end{equation}
where
\[
  J(\vc{q},\vc{p}) = \sum_{i=1}^{N-1} J_i =
  \sum_{i=1}^{N-1}\frac{\omega^2}{2} (q_i-q_{i+1})(p_i+p_{i+1}).
\]
The expectation value in (\ref{eq:defCJ}) refers to the stochastic time
evolution defined in section \ref{sec:dynamics} when the initial values
$(\vc{q}(0),\vc{p}(0))$ are distributed according to the equilibrium Gibbs
measure at temperature $T$.  The proof is a relatively tedious explicit
computation, which we do not report here in full detail.
\begin{proof} 
Define first the matrix $K$ by
\begin{equation}
  \label{eq:K}
  (K\vc{q})_i = \left\{ \begin{array}{ll} 
      q_{1}-q_{2}, & \text{for }i=1 \\
      q_{N-1}-q_{N}, & \text{for }i=N \\ 
      q_{i-1}-q_{i+1}, & \text{otherwise}
    \end{array} \right. ,
\end{equation}
so that
\[
 \frac{2}{\omega^2} J(\vc{q},\vc{p}) =
 \sum_{i=1}^{N-1}(q_i-q_{i+1})(p_i+p_{i+1}) = \vc{p}^T\! K\vc{q}.
\]
Then $\frac{1}{T^2} C_{JJ}^{(N)}(t;T) 
=\frac{\omega^4}{4}  g_N(t)$ for
\[
  g_N(t) = \frac{1}{T^2(N+1)} 
   \mean{\vc{p}(t)^T\! K\vc{q}(t)\, \vc{p}(0)^T\! K\vc{q}(0)}_{S_T}.
\]

The initial equilibrium measure is Gaussian with zero
mean and with a covariance $C(0,0) = T E$, where $E = T^{-1} \Seq $
is by (\ref{eq:eqlcov}) independent of $T$.  Correspondingly,
\[
 C(t,0) = \left\{ \begin{array}{ll}
     T \rme^{-t A} E, & \text{when }t \ge 0 \\ 
     T E \rme^{t A^T}, & \text{when }t<0
 \end{array}\right. .
\]
Applying the ``pairing rule'' of Gaussian correlations and setting
\[
 {\cal K} = \left( \begin{array}{cc}
     0 & K^T \\ K & 0
 \end{array} \right),  
\]
we then obtain, for $t\ge 0$,
\begin{equation}
  \label{eq:gNtrace}
  g_N(t) = \frac{1}{2(N+1)} 
    \tr\!\left[{\cal K}\rme^{-t A} E {\cal K}E \rme^{-t A^T}\right],
\end{equation}
and, for $t<0$, $g_N(t)=g_N(|t|)$.

Let us proceed by assuming the
existence of $\lim_{N \to \infty} g_N(t)$ and later  
comment on how to prove this.  From (\ref{eq:gNtrace}) we get
\begin{equation}
  \label{eq:gNdombound}
  |g_N(t)| \le \frac{\tr \, \1}{2(N+1)} 
    \norm{\rme^{-t A}}^2 \norm{{\cal K}}^2 \norm{E}^2.
\end{equation}
Since $\norm{\Phi^{-1}}= \sup_k 1/\mu_k \le 1/\gamma^2$, the norm of
$E$ is bounded uniformly in $N\to\infty$.  The same is clearly true
for $\norm{\cal K}$, and by (\ref{eq:exptAbound}) and (\ref{eq:gNdombound})
we can now apply dominated convergence in (\ref{eq:defKG}).  This proves
the integrability of the limit function, and yields
\begin{equation}
  \label{eq:defKG3}
 \kappa_{\rm GK} = \frac{\omega^4}{8} \lim_{N\to\infty} \frac{1}{N+1}
 \tr\!\left[{\cal K} \int_0^\infty\! \rmd t\, \rme^{-t A} E {\cal K}E
    \rme^{-t A^T}\right].
\end{equation}

We have now proved the first two equalities of the theorem.  We note
that the above argument, which allows to take the thermodynamic limit
out of the time-integral, would fail if $\gamma=0$, as then neither
the bound on $\norm{E}$ nor the exponential decay of 
$\norm{\rme^{-t A}}$ would be uniform in $N$.

Let us then denote
\[
  S' = \int_0^\infty\! \rmd t\, \rme^{-t A} E {\cal K}E \rme^{-t A^T} =
  \left( \begin{array}{cc}
        U' & Z' \\ (Z')^T & V'
      \end{array} \right)
\]
which is possible, as the integral clearly yields a symmetric operator.
Then
\begin{equation}
  \label{eq:RStrace}
 \tr\left[{\cal K}S'\right] = \tr\left[K^T (Z')^T + KZ'\right] = 
   2 \tr\bigl[\tilde{K}\tilde{Z}'\bigr]
\end{equation}
where $\tilde{K} = F^T K F$ and $\tilde{Z}'= F^T Z' F$.
Like the matrix $S$ defined by (\ref{e:defstationary}),
$S'$ is the unique solution of the equation
\[
  A S' + S'\! A^T = E {\cal K}E = \left( \begin{array}{cc}
      0 & \Phi^{-1} K^T \\ (\Phi^{-1} K^T)^T & 0
    \end{array} \right).
\]

In appendix \ref{sec:statstate} we prove that 
\[
   \tilde{Z}'_{kl} = -\frac{\lambda}{\omega^4} \frac{1}{G(c_k,c_l)}
  (\tilde{K}_-)_{kl},
\]
where $\tilde{K}_-$ is the antisymmetric part of $\tilde{K}$, and $G$ and
$c_k$ were defined in section \ref{sec:dynamics}. 
In particular, $\tilde{Z}'$ is antisymmetric, and thus by
(\ref{eq:RStrace}), 
\begin{equation}
  \label{eq:RSfinal}
 \tr\left[{\cal K}S'\right] = 2 \tr\bigl[ \tilde{K}_- \tilde{Z}' \bigr] =
  \frac{2 \lambda}{\omega^4} \sum_{k,l=1}^N \frac{1}{G(c_k,c_l)}
  (\tilde{K}_-)_{kl}^2 .
\end{equation}

Applying the definitions of $F$ and $K$ and neglecting all symmetric terms,
we get after some algebra
\[
  (\tilde{K}_-)_{kl} = -\frac{2}{N+1} \delta_{k-l,\rm odd}
  \frac{\sin\!\left(\frac{\pi k}{N+1}\right) 
    \sin\!\left(\frac{\pi l}{N+1}\right)}{
      \sin\!\left(\frac{\pi (k-l)}{2(N+1)}\right)
      \sin\!\left(\frac{\pi (k+l)}{2(N+1)}\right)}
\]
where $\delta_{u,\rm odd}=1$, if is $u$ is odd, and zero, if $u$ is even.
Observe then that for $l\approx k$ we have
\[
  (\tilde{K}_-)_{kl} \approx -\frac{4}{\pi (k-l)} \delta_{k-l,\rm odd}
  \sin\!\left(\frac{\pi k}{N+1}\right),
\]
while elsewhere $(\tilde{K}_-)_{kl} = \order{N^{-1}}$.  
Using this observation and the equality $\sum_{u\in\Z}
\delta_{u,\rm odd}/u^2 = \pi^2/4$, it is possible to prove that  
\[
  \frac{1}{N+1} \tr\left[{\cal K}S'\right] = 
    \frac{8 \lambda}{\omega^4} \frac{1}{N+1} 
    \sum_{k=1}^N \frac{\omega^2}{\lambda^2}
    \frac{ \sin^2\!\left(\frac{\pi k}{N+1}\right)}{
      \nu^2+4 \sin^2\!\left(\frac{\pi k}{2(N+1)}\right)} 
    + O\Bigl(\frac{1}{N}\Bigr).
\]

The same methods can be employed to  show that $\lim_{N\to\infty} g_N(t)$
exists for all $t>0$.   First write the trace in equation
(\ref{eq:gNtrace}) in the eigenspace of the force-matrix $\Phi$, and then
apply the above approximation to  $\tilde{K}_{kl}$ to find that only terms
with $k\approx l$  contribute and the contribution has a finite limit.

Combining the above with equation (\ref{eq:defKG3}), we have now proven that
\[
 \kappa_{\rm GK} = \frac{\omega^2}{\lambda} \int_0^1\!\!\rmd x \,
 \frac{\sin^2(\pi x)}{ \nu^2+ 4 \sin^2\bigl(\frac{\pi x}{2}\bigr)}
\]
which shows that $\kappa_{\rm GK}=\kappa$ for all $T$.
\end{proof}

\section{Non-uniform heat bath coupling}
\label{sec:gener}

Let us now consider the case when the heat bath couplings $\lambda_i$ are
not all equal and define  $\Lambda_{ij}= \delta_{ij} \lambda_i$.  As long
as $\norm{\rme^{-t A}}^2$ remains integrable, we can repeat the
computations in section \ref{sec:dynamics} and conclude that equations
(\ref{eq:sseqns2}) for the stationary covariance matrix  are still
valid. In particular, the matrix $Z$ is then antisymmetric.  Therefore, by
redoing the computations  in section \ref{sec:current}, we get the average
of  the energy transfer $R_i$ and of the current $J_i$ in the steady state
from the equations
\[
  \langle J_i \rangle_S = \omega^2 Z_{i,i+1} \qand \langle R_i \rangle_S =
   \lambda_i \left( T_i - V_{ii} \right).
\]

Thus the self-consistency condition still has the same form as before but,
as the earlier explicit solution of the steady state covariance $S$ is no
longer  possible, redoing the existence, uniqueness and local thermal
equilibrium results is not straightforward.  On physical grounds, we expect
the results to remain valid whenever there is an $N$-independent 
$\lambda>0$, such that the number of $i$'s for which 
$\lambda_i \geq \lambda$ is
proportional to $N$, certainly whenever this is true for all $i$.  Instead
of trying to redo the proofs,   we shall check what happens if we assume
that these results hold also when the  $\lambda_i$ are not all equal.

The equations for the stationary covariance
(\ref{eq:sseqns2}) now yield the relations
\begin{align}
  2 \mean{R_i}_S & = 2 \omega^2 (Z_{i-1,i} - Z_{i,i+1} ),  \\
 (\lambda_i+\lambda_{i+1}) Z_{i,i+1} & = \omega^2 (
  U_{ii} + U_{i,i+2} -  U_{i-1,i+1} -  U_{i+1,i+1} ). \label{eq:nonunZU}
\end{align}
The first equation implies that the current in the self-consistent 
steady state is constant, and then, by summing (\ref{eq:nonunZU}) over 
$i=1,\ldots,N-1$, we get
\begin{equation}\label{eq:nonuncurrent}
 \frac{2(N-1)\barlan }{ \omega^4} \JN =U_{11} - U_{NN} 
\end{equation}
where
\begin{equation}
  \label{eq:defbarlan}
\barlan =\frac{1}{N-1}\Bigl(
\sum_{i=1}^{N}\lambda_i - \frac{\lambda_N+\lambda_1}{2}\Bigr).  
\end{equation}

Let us next assume that the local equilibrium result proved in section
\ref{sec:localeql} is still valid, i.e.\ that for every $i,j$
\[
  U_{ij} = T_i \Phi^{-1}_{ij} + \order{\vep_N},
\]
where $\vep_N$ is defined by equation (\ref{eq:grad}), and that 
$\vep_N\to 0$  when $N\to\infty$.  Choosing the $\lambda_i$ such that
$\lim_{N\to\infty} \barlan = \bar\lambda > 0$, we get from
(\ref{eq:nonuncurrent}) the scaling of the total current,
\begin{equation}
  \label{eq:nonunlimJN}
\lim_{N\to\infty} (N-1) \JN = \bar\kappa (T_L-T_R)
\end{equation}
where $\bar\kappa$ is given by (\ref{eq:defkappa}), with  $\bar\lambda$
replacing $\lambda$ in the equation.

The conductivity will in general be space-dependent for non-uniform
couplings.  Consider, for instance, any sequence of couplings
$\Nlambda_{i}\ge 0$ which is bounded  (i.e.\ 
$\sup_{i,N} \Nlambda_{i}<\infty$) and for which the limit
\begin{equation} \label{eq:defLambda} 
  \Lambda(x)=\lim_{N\to\infty} \frac{1}{N} 
  \sum_{1\leq i\leq N x}\Nlambda_{i}
\end{equation}
exists for all $x\in[0,1]$ and defines a smooth function $\Lambda$ with
$\Lambda(1)>0$.

By summing (\ref{eq:nonunZU}) over a range of indices from $1$ to $j-1$ we
get, after applying the local equilibrium assumption and (\ref{5.22}), that
\[
 \frac{2\JN}{\omega^4 \Phi^{-1}_{11}} \Bigl( \sum_{i=1}^{j}\Nlambda_i -
  \frac{\Nlambda_{j}+\Nlambda_{1}}{2}\Bigr) 
  = T_{L} -  T_{j} + \order{\vep_N}.
\]
Then, by applying (\ref{eq:nonuncurrent}) and (\ref{eq:defLambda}), we can
conclude that the temperature profile now converges to
\[
  T(x) = T_L - (T_L-T_R) \frac{\Lambda(x)}{\Lambda(1)},
\]
and, therefore, that Fourier's law is satisfied  with a thermal conductivity
\begin{equation}
  \label{eq:defkappax}
  \kappa(x) =  \frac{\omega^2}{\lambda(x)}  
  \frac{1}{2+\nu^2+\sqrt{\nu^2 (4+\nu^2)} }
\end{equation}
where $\lambda(x)=\frac{\rmd}{\rmd x}\Lambda(x)$.  Note  that on any
interval on which $\lambda(x)=0$ the local conductivity is infinite and the
temperature profile remains constant.

For instance, if $\lambda_i=\lambda$ for every $m$:th coupling and
$\lambda_i=0$ otherwise, we get a finite conductivity equal to $m$ times
the one computed in section \ref{sec:fourier}.  Taking $m=N$ we then
(formally) recover the linear divergence of the conductivity in $N$ which
was found in \cite{lebo67}.

\section{Higher dimensions}
\label{sec:crystal}

Here we extend the results of the previous sections to a system of
oscillators first in $d = 2$ and then also in higher dimensions. The
solution can be obtained in a way very similar to what we did in sections
\ref{sec:dynamics} to \ref{sec:greenkubo}, and  we shall just report the
necessary adjustments. Moreover, we shall only consider explicitly the
system in two dimensions. No real modifications are necessary to extend the
computations to higher dimensions.

The Hamiltonian for the system is now given by
\begin{align}
  \label{eq:hamiltonian2}
  H(\vc{q},\vc{p}) = & \sum_{i=1}^N \sum_{j=1}^{N'} \Bigl[\frac{1}{2}
    p_{i,j}^2 + u(q_{i,j})\Bigr] \nonumber\\ 
    & + \sum_{j=1}^{N'} \sum_{i=1}^{N+1} v(q_{i,j}-q_{i-1,j})+ 
    \sum_{i=1}^{N} \sum_{j=1}^{N'} v(q_{i,j}-q_{i,j-1})
\end{align}
where we assume, as before, that $q_{0,j}=q_{N+1,j}=0$ and we fix periodic
boundary conditions in the second direction, i.e.\ $q_{i,0}=q_{i,N'}$.  As
in section \ref{sec:dynamics}, the oscillator at  the site $(i,j)$ is
coupled to a Langevin heat bath with temperature $T_{i,j}$, and we set
$T_{1,j} = T_L$, $T_{N,j} = T_R$ while all other $T_{i,j}$ are to be
determined self-consistently. Thus the time evolution is still defined by
equations (\ref{e:OUprocess})--(\ref{eq:defphi}) if we interpret the
operator $\Delta$ in (\ref{eq:defphi}) as the discrete Laplacian in two
dimensions with mixed boundary conditions: Dirichlet in the first direction
and periodic in the second direction.

Extending the results in section \ref{sec:current}, we first define the
local energy by
\begin{align}
H_{i,j}(\vc{q},\vc{p}) =& \frac{1}{2}p_{i,j}^2+u(q_{i,j})
 + \frac{1}{2}[v(q_{i,j}-q_{i-1,j})+v(q_{i+1,j}-q_{i,j})\nonumber \\
 & + v(q_{i,j}-q_{i,j-1})+v(q_{i,j}-q_{i,j+1})]
\end{align}
again  with double contribution for the terms involving $q_{0,j}$ and
$q_{N+1,j}$.  Then the analog of (\ref{eq:energyflux}) is true if we 
define the current as a two dimensional vector with 
$J^1_{i,j} = 0$ for $i=0,N$, and with the other components given by
\begin{align}
J_{i,j}^1(\vc{q},\vc{p})&=-\frac{\omega^2}{2}(q_{i+1,j}-q_{i,j})
(p_{i,j}+p_{i+1,j}), \\
J_{i,j}^2(\vc{q},\vc{p})&=-\frac{\omega^2}{2}(q_{i,j+1}-q_{i,j})
(p_{i,j}+p_{i,j+1}),
\end{align}
where $q_{i,j}$ and $p_{i,j}$ are periodic in $j$.

The source terms are given by $R_{i,j} = \lambda (T_{i,j} - (p_{i,j})^2)$,
and the self-consistency condition thus becomes
\[
T_{i,j}=\langle (p_{i,j})^2\rangle_S,\quad \text{for }i=2,\ldots,N-1
\text{ and }j=1,\ldots,{N'}
\]
with $T_{1,j}=T_L$ and $T_{N,j}=T_R$.  The main observation is that, as in
\cite{nakazawa70},  we can Fourier transform this system in the periodic
direction and obtain a system of decoupled chains.

More precisely, let for $k=1,\ldots,{N'}$
\begin{equation}
   q_{i}(k) =\frac{1}{\sqrt{N'}}\sum_{j=1}^{N'}  q_{i,j}
   \rme^{\ci\frac{2 \pi}{N'}k j}\quad\text{when}\quad
   q_{i,j}=\frac{1}{\sqrt{N'}}\sum_{k=1}^{N'}
   q_{i}(k)\rme^{-\ci\frac{2 \pi}{N'}k j},
\end{equation}
and define $p_{i}(k)$ analogously.  This corresponds to a change to a
(complex) eigenbasis of the periodic Laplacian, and we obtain that, for any
fixed $k$, $\vc{q}(k)$ and $\vc{p}(k)$ satisfy equation (\ref{e:OUprocess})
with the only difference that now the potential $\Phi$ is given by
(\ref{eq:defphi}) with $\nu^2$ replaced by
$\nu(k)^2=\nu^2+2(1-\cos(\frac{2\pi k}{N'}))\ge \nu^2$.   However, the
noise term will then become more complicated and it can still  
{\it a priori\/} couple the components with different values of $k$.

In general, $\vc{p}(k)$ and $\vc{q}(k)$ are complex numbers, and the
stochastic equations should be understood applying to the real and
imaginary part separately.  However, as $A$ remains a real matrix, equation
(\ref{eq:Cteq}) still holds if we replace the matrix $\Sigma^2$ by
\begin{equation}
  \label{eq:complexsigma}
\begin{pmatrix} 0 & 0 \\ 0 & \sigma\sigma^\dagger \end{pmatrix}, 
\text{ where } (\sigma\sigma^\dagger)_{i,k;i',k'} =  \delta_{ii'} 2\lambda
\frac{1}{N'} \sum_j T_{i,j} \,\rme^{\ci\frac{2 \pi}{N'} j (k-k')}.
\end{equation}
On the other hand, our bound for the norm of the exponential of $A$  is
obviously still valid, and we can conclude that for every temperature
profile there is a unique stationary state which is reached exponentially
fast and which is determined by equation (\ref{e:defstationary})  with the
matrix (\ref{eq:complexsigma}) replacing  $\Sigma^2$ there.

Next we need to prove the existence and uniqueness of the self-consistent
temperature profile.  In fact, Theorem \ref{th:scprof} is valid also in the
higher dimensional case considered here, but since the proof remains
essentially unchanged, we do not include it here.

The boundary conditions we impose are constant in the periodic direction,
and we expect from symmetry that the self-consistent temperature profile is
also constant in that  direction, even for finite $N$.  This is also
directly implied by  the above quoted uniqueness since, if $(t_{i,j})$ is a
self-consistent profile, then also its translates, 
$T_{i,j}= t_{i,j+j'}$ for any $j'$, are 
self-consistent with the same boundary conditions.

Consider thus a temperature profile $T_{i,j} = \tau_i$ for which
$\tau_1=T_L$ and $\tau_N=T_R$.  By (\ref{eq:complexsigma}), we then have
always
\[
  (\sigma\sigma^\dagger)_{i,k;i',k'} = \delta_{ii'} \delta_{kk'} 
  2\lambda \tau_i . 
\]
Applying this in the equation corresponding to (\ref{e:statstate})
reveals that the components having different values of $k$ then
become independent in the steady state.  In particular,
\[
  \langle p_i(k)p_j(k')^* \rangle_S=\langle
  p_i(k)p_j(k)^*\rangle_S\,\delta_{kk'} ,
\]
and, therefore for all $i,j$,
\begin{equation}
  \label{eq:ppcorr}
  \mean{p_{i,j}p_{i,j}}_S = \frac{1}{N'} \sum_{k,k'=1}^{N'}   
  \rme^{-\ci\frac{2 \pi}{N'}j (k-k')} \mean{ p_i(k)p_i(k')^* }_S
  =  \frac{1}{N'} \sum_{k=1}^{N'} \mean{ p_i(k)p_i(k)^* }_S.  
\end{equation}
Here the expectation value can be computed by
$\mean{ p_i(k)p_i(k)^* }_S = (M(k)\vc{\tau})_i$,
where  $M(k)=\left. M\right|_{\nu^2=\nu(k)^2}$ and $M$ is the matrix
defined in section \ref{sec:selfcons}.  Therefore, simply by replacing the
matrix $M$ with $(N')^{-1}\sum_k M(k)$ we can repeat the computations in
section \ref{sec:selfcons}, and find a vector $\vc{\tau}$ which leads to a
self-consistent profile $T_{i,j}$.

It is then easy to see, applying  the decoupling of the modes as above and
then using the antisymmetry of the covariance component $Z$, that there is
no average current in the second direction, i.e.\ 
$\langle J_{i,j}^2\rangle_S=0$.  Similarly, we get 
for all $i=1,\ldots,N-1$  the result
\begin{equation}
  \mean{J_{i,j}^1}_S = \omega^2 Z_{i,j;i+1,j} = 
  \frac{1}{N'} \sum_{k=1}^{N'} \left.\JN\right|_{\nu^2=\nu(k)^2}
\end{equation}
where $\JN$ denotes the current through the corresponding chain.

Repeating the computations in section \ref{sec:fourier} and using the above
decoupling of the $k$-modes, we can then conclude that  local equilibrium
holds in the limit $N,N'\to\infty$ (for this one needs to notice that the
exponential decay of correlations is uniform in $k$, as
$\nu(k)^2\ge\nu^2>0$), the limiting temperature profile is constant in  the
periodic direction and connects $T_L$ and $T_R$ linearly in the first
direction.  Fourier's law is also satisfied with  the conductivity now
given by
\begin{equation}
\kappa=\lim_{N'\to\infty}\frac{1}{N'}\sum_k\kappa(k)
=\frac{\omega^2}{\lambda}\int_0^1\!
\frac{\rmd y}{2+\tilde\nu(y)^2+\sqrt{\tilde\nu(y)^2 (4+\tilde\nu(y)^2)}}
\end{equation}
where $\tilde\nu(y)^2=\nu^2+2(1-\cos(2 \pi y))$. 

For the system with $d-1$
extra periodic dimensions, we could analogously arrive at the same
conclusions, but with a conductivity 
\begin{equation}\label{eq:7.10}
\kappa=\frac{\omega^2}{\lambda}
\int_{[0,1]^{d-1}}
\frac{\rmd^{d-1} y}{2+\tilde\nu(\vc{y})^2+\sqrt{\tilde\nu(\vc{y})^2
    (4+\tilde\nu(\vc{y})^2)}} 
\end{equation}
where now $\tilde\nu(\vc{y})^2=\nu^2+2\sum_{i=1}^{d-1} (1-\cos(2 \pi y_i))$.
Observe, in particular, that the 
conductivity steadily decreases with each added dimension.  We prove in
appendix \ref{sec:asymptotics} that the 
asymptotic behavior of the conductivity in the limit $d\to\infty$
is given by
\[
  \kappa=\frac{\omega^2}{4 d \lambda} ( 1 + o(1) )
\]
where the correction term depends only on $\nu$ and $d$.  Thus by choosing
a suitable sequence of $\lambda = O(d^{-1})$, we can
have $\lambda \to 0$ when $d\to\infty$ and still keep the 
conductivity finite and constant.

It would also be straightforward to check that the Green-Kubo formula holds
in the higher dimensional case.  More precisely, Theorem \ref{th:GK} is
still valid for the above system in $d$ dimensions, if the current-current
correlator is defined instead of (\ref{eq:defCJ}) by
\[
  C_{JJ}^{(N)}(t;T) = \frac{1}{d\,\prod_i\! N_i}
  \mean{\vc{J}(\vc{q}(t),\vc{p}(t))\cdot\vc{J}(\vc{q}(0),\vc{p}(0))}_{S_T}.
\]
For proving this, the $\mean{J^1 J^1}$-term can be analyzed exactly as
before,  while the analysis of the remaining $\mean{J^i J^i}$-terms in  the
periodic directions will be even simpler, as in the complex eigenbasis used
here the operator corresponding to $K$ will be exactly diagonal.

\section{Discussion}

We raise again the question, discussed extensively in \cite{bonetto00}  and
\cite{lebo78}, of whether it is possible to derive Fourier's law for a
system with purely Hamiltonian bulk dynamics.  There are two ways of
formulating this problem:  ({\it i\/}) The system could be fully isolated
and evolving towards equilibrium from an initial nonuniform local
equilibrium state.  ({\it ii\/}) The system could be maintained in a
stationary non-equilibrium state by coupling it at the boundaries to
infinite  reservoirs, either stochastically as in \cite{lebo67} 
(or variations thereof, see \cite{bonetto00}) or mechanically as in
\cite{LS78,eckm99}.  One could also keep the  temperature fixed at the end
of the system by means of deterministic Gaussian thermostats
\cite{gallcoh95}.

In the first case this amounts to proving the existence of a hydrodynamical
scaling limit on the dissipative time scale.  This is a well known,
extremely difficult problem \cite{spohn:hydro}.  It is clearly not true for
the harmonic crystal or other integrable models but is believed to be true
for macroscopic systems with more realistic type of interactions, e.g.\
hard spheres or with  Lennard-Jones potentials.  For anharmonic crystals,
the kind considered in \cite{eckm99},  one would have to go beyond the
Kolmogorov, Arnold, Moser domain \cite{taborbook}  and presumably also
beyond the Fermi, Pasta, Ulam \cite{FPU55} models \cite{lepri01}.  The only
mechanical system, for which such a result has been derived, is for the
highly degenerate model of a macroscopic system of independent particles
moving in a periodic array of scatterers, i.e.\ for the multi-particle
Sinai billiard, where one proves Fick's law, the analog of Fourier's law
for the conserved particle current \cite{LS82}.

In the second case of stationary non-equilibrium states one may hope to
prove a global Fourier's law, i.e.\ we want 
${\cal L} J_{\cal L}/(T_L - T_R) \to \kappa$  
as ${\cal L} \to \infty$.  Here ${\cal L}$ is the distance, 
in microscopic units, between the boundaries of the system, say a
cylinder, maintained at fixed temperatures $T_L$ and $T_R$.  We want a
$\kappa$ which depends only on the bulk properties of the system.  We
expect further that when $T_L \to T_R = T$, the limit of $\kappa$ should
coincide with the heat conductivity $\kappa(T)$  at the local equilibrium
temperature $T$ in the isolated time-evolving case ({\it i\/}).  Again the
only mechanical system for which such a result has been proven is for the
degenerate system of point particles moving among a periodic array of
scatterers where the heat current is really a particle current (particles
pick up energy at the right wall) \cite{LS78}.  The best that has been
proven for other systems is the existence of a stationary state
\cite{eckm99, GLP81, GKI85} and the positivity of $(T_L - T_R)J_{\cal L}$
for fixed ${\cal L}$ \cite{lucr02b}.

The results proven in this paper make use of the stochastic interactions in
the bulk to produce a local equilibrium state.  This is in the spirit of
the general work in the last two decades proving the existence of
hydrodynamical laws in the appropriate scaling limits for systems evolving
via stochastic dynamics \cite{spohn:hydro}.  We should mention here in
particular the work of Kipnis, Marchioro and Presutti \cite{KMP82} who
proved results similar to ours for a model with purely stochastic internal
dynamics.  They were in fact able to consider a situation where the energy
is strictly conserved in the bulk rather than just in the average as in the
model considered here.  This can be done also for a modified (more
mechanical) version of their model considered by Olla \cite{ollanote} in
which there is an energy conserving Ornstein-Uhlenbeck type process
producing an energy exchange between neighboring oscillators.

The main advantage of the BRV self-consistent model is that the
average energy flow along the temperature gradient is, 
as seen in (\ref{eq:Jidef}), entirely Hamiltonian.  
As mentioned in the introduction, it might in fact
be possible to make our model entirely mechanical by coupling each site to
a large Hamiltonian reservoir, {\it a la}\/  Ford, Kac and Mazur
\cite{FKM65}, which would produce an effective stochastic reservoir that
would automatically, without any imposition of self-consistency, be at the
right temperature.  

This is in fact what seems to happen effectively when we let 
the dimension of the crystal go to infinity.  As shown in Appendix 
\ref{sec:asymptotics},
after taking the limits $t \to \infty$ and $N \to \infty$, we can
let the coupling to the interior heat baths, which we denote by
$\dlambda$, go to zero as $d^{-1}$, 
and still obtain a finite value of the conductivity.  It is clear from
the analysis in Section \ref{sec:gener} that, if we set 
$\lambda_1 = \lambda_N = \ell_0$ and 
$\lambda_2 = \lambda_3 = \cdots = \lambda_{N-1} = \dlambda$,
then the heat conductivity is obtained by replacing $\lambda$ by
$\dlambda$ in (\ref{eq:defkappa}) and in (\ref{eq:7.10}).

An open interesting problem is to consider our model for an anharmonic
crystal, e.g.\ by setting in (\ref{eq:defUV}),  
$u(q) = \frac{1}{2}\gamma^2 q^2 + \frac{1}{2} \delta q^4$.   
We expect that for a fixed $\delta > 0$
the heat conductivity $\kappa$ would have a finite limit as the auxiliary
couplings with the interior heat baths  are taken to zero. It might even be
possible to prove such a result by {\it starting}\/ with a perturbation
expansion in $\delta$ around the local equilibrium stationary state found
here and then doing a suitable resummation or applying a renormalization
group type argument.  See however, the results of the perturbation
expansion in the case with purely Hamiltonian bulk dynamics derived in
\cite{lefev03}.

We note finally that the harmonic heat conductivity $\kappa$ given in
(\ref{eq:defkappa}) would remain finite if we let $\lambda \to 0$ and
$\gamma \to \infty$ in such a way that 
$\lambda \gamma^2 \to \alpha > 0$. It is not clear whether 
this limit has any physical significance.

\section*{Acknowledgments}

We would like to thank the Institute for Advanced Study in Princeton,
New Jersey, USA, for generous hospitality making this project
possible.  We also want to thank Antti Kupiainen and Herbert Spohn
for helpful discussions.  J.\ Lukkarinen acknowledges the
financial support for this project by the Academy of Finland grants
Nr.\ 100438 and Nr.\ 200231.  This work was also supported by NSF
Grant DMR 01-279-26 and by AFOSR Grant 49620-01-1-0154.

\appendix 

\section{Bound for the time-evolution matrix}
\label{sec:boundexpA}

Let $F$ be the orthogonal matrix defined by equation (\ref{eq:defF}), and
define
\begin{equation}
  \label{eq:defcalF}
 {\cal F} = \left( \begin{array}{cc}
     F & 0 \\ 0 & F
 \end{array} \right). 
\end{equation}
As $F$ diagonalizes $\Phi$, we then easily see that 
$A = {\cal F} \tilde{A} {\cal F}^T$, 
where (with $Y$ again denoting the eigenvalue matrix of $\Phi$)
\[
  \tilde{A} = \left( \begin{array}{cc}
      0 & -\1 \\ Y & \lambda\1
 \end{array} \right).
\]

Since $\tilde{A}$ is block diagonal (after a permutation of indices) and
${\cal F}$ is orthogonal, it follows that the norm of the exponential
satisfies
\begin{equation}
  \label{eq:maxAk}
  \norm{\rme^{-t A}} = \max_{k} \norm{\rme^{-t A_k}}
\end{equation}
where for each $k$ we have defined
\begin{equation}
 A_k = \label{eq:atildek} \left( \begin{array}{cc}
     0 & -1 \\ \mu_k & \lambda
  \end{array} \right)  .
\end{equation}

The eigenvalues of $A_k$ are
\[
  \alpha_k^{\pm} = \frac{\lambda}{2} \pm \rho_k \quad\text{where}\quad
  \rho_k = \sqrt{\frac{\lambda^2}{4} - \mu_k},
\]
and it is easy to see that
\[
 \re\,\alpha_k^{\pm}\geq  \underline{\alpha} =
\min\Bigl\{\frac{\lambda}{2},\frac{\gamma^2}{\lambda} \Bigr\}>0. 
\]
However, since $A_k$ is not symmetric (in fact, there are values of the
parameters when it is not even diagonalizable) we have to take more care in 
analyzing the norm of its exponential.  Performing the Jordan decomposition
of $A_k$ explicitly yields
\begin{equation}
   \rme^{-t A_k} = \rme^{-t \lambda/2} \cosh(\rho_k t)
   \Bigl[\, \1 + \frac{\tanh(\rho_k t)}{\rho_k}
   \begin{pmatrix}
       \lambda/2 & 1 \\ -\mu_k & -\lambda/2
   \end{pmatrix}
   \Bigr] 
\end{equation}
from which we straightforwardly arrive at the following
bound valid for $t\ge 0$,
\begin{equation*}
    \left\Vert \rme^{-t A_k}\right\Vert \le \rme^{-t \underline{\alpha} }
    \left[ 1 + t (1+\gamma^2+ 4 \omega^2+ \lambda/2) \right].
\end{equation*}

Applying this to (\ref{eq:maxAk}) easily yields the conclusion in section 
\ref{sec:dynamics}, at equation (\ref{eq:exptAbound}).
We remark that if $\gamma=0$, we could still have a lower bound
$\underline{\alpha}>0$, but it would not be uniform in $N$.  In fact, since
then $\inf_k \mu_k = \order{N^{-2}}$, we would then need to take also
$\underline{\alpha}=\order{N^{-2}}$.

\section{Solution of the stationary covariance}
\label{sec:statstate}

We derive here an explicit solution to the equation
\begin{equation}
  \label{eq:assab}
  A S + S A^T = \Sigma^2,
\end{equation}
which---for the matrix $\Sigma^2$ used in section \ref{sec:dynamics}---will
yield the stationary covariance matrix.   The matrix
$A$ is defined as in (\ref{eq:defA}), but we need 
the solution for a more general ``noise matrix'' in section
\ref{sec:greenkubo}. Therefore, we consider here 
\[
  \Sigma^2 = \left( \begin{array}{cc}
      0 & b \\ b^T & 2 \lambda d
 \end{array} \right)
\]
where $b$ and $d$ are real $N\times N$ matrices and $d^T = d$. 

Denoting
\[
  S = \left( \begin{array}{cc}
      U & Z \\ Z^T & V
 \end{array} \right),
\]
we get that $S$ is a solution to (\ref{eq:assab}) if and only if its
components satisfy
\begin{align}\label{eq:sseqns}
\begin{split}
  Z^T & = -Z \\
  V &  = \frac{1}{2} ( \Phi U + U \Phi ) -b_+ \\ 
  \lambda Z & = \frac{1}{2} ( \Phi U - U \Phi) + b_-  \\
  2 \lambda ( d + b_+ ) & = \Phi Z - Z \Phi + \lambda ( U\Phi + \Phi U )
\end{split}
\end{align}
where $b_+$ and $b_-$ are the symmetric and the antisymmetric part of $b$.

If we define
\[
   \tilde{d} = F^T\! d F \qand \tilde{V} = F^T V\! F,
\]
and also $\tilde{b}$, $\tilde{U}$ and $\tilde{Z}$ similarly, then we get
\begin{align*}
  \tilde{U}_{kl} & = \frac{2}{g_{kl}} \left[ 2 \lambda^2
    ( \tilde{d}_{kl} + (\tilde{b}_+)_{kl} ) - 
    (\mu_k-\mu_l)  (\tilde{b}_-)_{kl} \right] \\
  \tilde{V}_{kl} & = \frac{1}{g_{kl}} \left[ 2 \lambda^2 ( \mu_k + \mu_l )
    \tilde{d}_{kl} - (\mu_k-\mu_l) 
    \left(\mu_k \tilde{b}_{kl} - \mu_l \tilde{b}_{lk}\right) \right] \\
  \tilde{Z}_{kl} & = \frac{2\lambda}{g_{kl}} \left[ 
    (\mu_k-\mu_l) \tilde{d}_{kl} + 
    \mu_k \tilde{b}_{kl} - \mu_l \tilde{b}_{lk} \right]
\end{align*}
where $\mu_k$ are the eigenvalues of $\Phi$, and for all $k$ and $l$
\[
  g_{kl} = 4 \omega^4 G(c_k,c_l)>0
\]
where $c_k$ and $G$ are defined in equations (\ref{eq:defck}) and
(\ref{eq:defG}), respectively.

When $b=0$ and $d_{ij}=\delta_{ij} T_i$ we get
(\ref{eq:Bij})--(\ref{eq:deffX}).   In section \ref{sec:greenkubo} we need
to know $\tilde{Z}_{kl}$ when $d=0$ and $b=\Phi^{-1}K^T$ with $K$ defined
by (\ref{eq:K}).  Since then $\tilde{b}_{kl} = \tilde{K}_{lk}/\mu_k$, where
$\tilde{K}= F^T K F$, we get
\[
  \tilde{Z}_{kl} = -\frac{\lambda}{\omega^4} \frac{1}{G(c_k,c_l)}
  (\tilde{K}_-)_{kl}
\]
with $\tilde{K}_-$ denoting the antisymmetric part of $\tilde{K}$.

\section{Asymptotic behavior of the conductivity}
\label{sec:asymptotics}

In section \ref{sec:crystal} we derived a formula for the conductivity of
the $d$-dimensional crystal,
\[
\kappa=\frac{\omega^2}{\lambda} I\quad\text{ where }\quad
I=\int_{[0,1]^{d-1}}
\frac{\rmd^{d-1} y}{2+\tilde\nu(\vc{y})^2+
\sqrt{\tilde\nu(\vc{y})^2 (4+\tilde\nu(\vc{y})^2)}}
\]
and $\tilde\nu(\vc{y})^2=\nu^2+2\sum_{i=1}^{d-1} (1-\cos(2 \pi y_i))$.
Here we prove that the asymptotic behavior of $I$ for $d\to\infty$ is
given for any fixed $\nu> 0$ by
\begin{equation}
  \label{eq:Iasymp}
 I = \frac{1}{4 d} ( 1 + o(1) ).
\end{equation}

First we point out that for all $r\ge 0$
\[
  \frac{1}{2+r^2+\sqrt{r^2 ( 4+r^2 )} } = \int_0^1\!\!\rmd x \,
  \frac{\sin^2(\pi x)}{ r^2+ 4 \sin^2\bigl(\frac{\pi x}{2}\bigr)}  =
  \int_0^1\!\!\rmd x \, 
  \frac{\sin^2(2 \pi x)}{ r^2+ 4 \sin^2\bigl(\pi x\bigr)}
\]
and, therefore,
\[
I=\int_{[0,1]^{d}}\!\! \rmd^{d} y 
\frac{\sin^2(2 \pi y_1)}{ \nu^2+ 4\sum_{i=1}^d \sin^2\bigl(\pi y_i\bigr)}.
\]
Since the denominator is always strictly positive, we can then use the
formula $1/r = \int_0^\infty\! \rmd t\, \exp(-t r)$ valid for all $r>0$ and
obtain
\[
I = \int_0^\infty\! \rmd t\, \rme^{-t \nu^2} I_1(t) I_0(t)^{d-1}
\]
where
\begin{align*}
  I_0(t) & = \int_0^1\!\rmd y\, \rme^{-4 t \sin^2(\pi y)},
  \quad\text{ and}\\ 
  I_1(t) & = \int_0^1\!\rmd y\, \sin^2(2\pi y) \rme^{-4 t \sin^2(\pi y)}.
\end{align*}

Now both functions $I_i(t)$, $i=0,1$, are clearly continuous and  strictly
monotonously decreasing from $I_i(0)$ to $0$ when $t$ goes from $0$ to
$\infty$, with $I_0(0)=1$ and $I_1(0)=\frac{1}{2}$.   In addition, $I_0$ is
bounded for all $t\ge 0$ by
\begin{equation}
  \label{eq:I0bound}
  I_0(t) \le \frac{1}{\sqrt{1+t}}.
\end{equation}
This follows from
\begin{align*}
 I_0(t) & = \int_0^1\!\rmd x\, \rme^{-4 t \sin^2(\pi x/2)} 
  \\ & \le \int_0^1\!\rmd x\, \rme^{-4 t x^2} = \frac{1}{\sqrt{1+t}}  
 \int_0^{\sqrt{1+t}}\!\rmd y\,\rme^{-4 y^2 t/(1+t)}
\end{align*}
since the derivative of the
last integral is negative for $t\ge 0$.

By dominated convergence we then find that, when $d\to\infty$,
\[
  d \int_1^\infty\! \rmd t\, \rme^{-t \nu^2} I_1(t) I_0(t)^{d-1} \to 0.
\]
Changing variables to $s=t d$ in the remaining integral shows then that
\begin{equation}
  \label{eq:dI}
  d I = \int_0^{d}\! \rmd s\, \rme^{-\nu^2 s/d} I_1\Bigl(\frac{s}{d}\Bigr)
  I_0\Bigl(\frac{s}{d}\Bigr)^{d-1} + o(1).
\end{equation}
Since $1+x\ge 2^x$ for all $0\le x\le 1$, inequality (\ref{eq:I0bound})
yields the bound
\[
 I_0\Bigl(\frac{s}{d}\Bigr)^{d-1} \le 2^{-s/4}
\]
for all $0\le s\le d$ and $d\ge 2$.  This means that dominated convergence
can also be applied to the integral in (\ref{eq:dI}), and as 
$I_0(s/d) = 1- 2 s /d + O(d^{-2})$, we then find
\[
  \lim_{d\to\infty} (d I) = \frac{1}{2} \int_0^\infty\! \rmd s\, 
  \rme^{-2 s} =  \frac{1}{4}
\]
which proves the equation (\ref{eq:Iasymp}).


\begin{thebibliography}{10}

\bibitem{bonetto00}
F.~Bonetto, J.~L. Lebowitz, and L.~Rey-Bellet,
{\it {F}ourier's law: a challenge to theorists\/}.
\newblock In A.~Fokas, A.~Grigoryan, T.~Kibble, and B.~Zegarlinski (eds.), {\it
  Mathematical Physics 2000\/}, pp. 128--150, London, 2000. Imperial College
  Press.

\bibitem{lepri01}
S.~Lepri, R.~Livi, and A.~Politi,
{\it Thermal conduction in classical low-dimensional lattices\/},
Phys. Rep.~{\bf 377} (2003) 1--80.

\bibitem{bolrich70}
M.~Bolsterli, M.~Rich, and W.~M. Visscher,
{\it Simulation of nonharmonic interactions in a crystal by self-consistent
  reservoirs\/},
Phys. Rev.~{\bf A 4} (1970) 1086--1088.

\bibitem{rich75}
M.~Rich and W.~M. Visscher,
{\it Disordered harmonic chain with self-consistent reservoirs\/},
Phys. Rev.~{\bf B 11} (1975) 2164--2170.

\bibitem{lebo67}
Z.~Rieder, J.~L. Lebowitz, and E.~Lieb,
{\it Properties of a harmonic crystal in a stationary nonequilibrium state\/},
J. Math. Phys.~{\bf 8} (1967) 1073--1078.

\bibitem{nakazawa70}
H.~Nakazawa,
{\it On the lattice thermal conduction\/},
Suppl. Progr. Theor. Phys.~{\bf 45} (1970) 231--262.

\bibitem{spohn:hydro}
H.~Spohn,
{\it Large Scale Dynamics of Interacting Particles\/}.
\newblock Springer, Berlin, 1991.

\bibitem{oksendal}
B.~{\O}ksendal,
{\it Stochastic differential equations: an introduction with applications\/}.
\newblock Springer, Berlin, fifth edition, 1998.

\bibitem{lebspohn99}
J.~L. Lebowitz and H.~Spohn,
{\it A {G}allavotti-{C}ohen type symmetry in the large deviation functional for
  stochastic dynamics\/},
J. Stat. Phys.~{\bf 95} (1999) 333--365.

\bibitem{eyink96}
G.~L. Eyink, J.~L. Lebowitz, and H.~Spohn,
{\it Hydrodynamics and fluctuations outside of local equilibrium: Driven
  diffusive systems\/},
J. Stat. Phys.~{\bf 83} (1996) 385--472.

\bibitem{lebo78}
J.~L. Lebowitz,
{\it Exact results in nonequilibrium statistical mechanics: Where do we
  stand?\/},
Suppl. Progr. Theor. Phys.~{\bf 64} (1979) 35--49.

\bibitem{LS78}
J.~L. Lebowitz and H.~Spohn,
{\it Transport properties of the {L}orentz gas: {F}ourier's law\/},
J. Stat. Phys.~{\bf 19} (1978) 633--654.

\bibitem{eckm99}
J.-P. Eckmann, C.-A. Pillet, and L.~Rey-Bellet,
{\it Non-equilibrium statistical mechanics of anharmonic chains coupled to two
  heat baths at different temperatures\/},
Commun. Math. Phys.~{\bf 201} (1999) 657--697.

\bibitem{gallcoh95}
G.~Gallavotti and E.~G.~D. Cohen,
{\it Dynamical ensembles in stationary states\/},
J. Stat. Phys.~{\bf 80} (1995) 931--970.

\bibitem{taborbook}
M.~Tabor,
{\it Chaos and Integrability in Nonlinear Dynamics: An Introduction\/}.
\newblock Wiley, New York, 1989.

\bibitem{FPU55}
E.~Fermi, J.~Pasta, and S.~Ulam,
{\it Studies in nonlinear problems, {I}\/}.
\newblock In A.~C. Newell (editor), {\it Nonlinear Wave Motion\/}, pp.
  143--156. American Mathematical Society, Providence, RI, 1974.
\newblock Originally published as Los Alamos Report LA-1940 in 1955.

\bibitem{LS82}
J.~L. Lebowitz and H.~Spohn,
{\it Microscopic basis for {F}ick's law of self-diffusion\/},
J. Stat. Phys.~{\bf 28} (1982) 539--556.

\bibitem{GLP81}
S.~Goldstein, J.~L. Lebowitz, and E.~Presutti,
{\it Stationary states for a mechanical system with stochastic boundaries\/}.
\newblock In J.~Fritz, J.~L. Lebowitz, and D.~Sz\'{a}sz (eds.), {\it Random
  Fields (Colloquia Mathematicae Societatis J\'{a}nos Bolyai 27)\/}, pp.
  403--419, Amsterdam, 1981. North-Holland.

\bibitem{GKI85}
S.~Goldstein, C.~Kipnis, and N.~Ianiro,
{\it Stationary states for a system with stochastic boundary conditions\/},
J. Stat. Phys.~{\bf 41} (1985) 915--939.

\bibitem{lucr02b}
L.~Rey-Bellet and L.~E. Thomas,
{\it Fluctuations of the entropy production in anharmonic chains\/},
Ann. H. Poinc.~{\bf 3} (2002) 483--502.

\bibitem{KMP82}
C.~Kipnis, C.~Marchioro, and E.~Presutti,
{\it Heat flow in an exactly solvable model\/},
J. Stat. Phys.~{\bf 27} (1982) 65--74.

\bibitem{ollanote}
S.~Olla.
\newblock Private communication.

\bibitem{FKM65}
G.~W. Ford, M.~Kac, and P.~Mazur,
{\it Statistical mechanics of assemblies of coupled oscillators\/},
J. Math. Phys.~{\bf 6} (1965) 504--515.

\bibitem{lefev03}
R.~Lefevere and A.~Schenkel,
{\it Perturbative analysis of anharmonic chains of oscillators out of
  equilibrium\/},
preprint (2003), {\tt http://arxiv.org/abs/math-ph/0303050}.

\end{thebibliography}

\end{document}